\documentclass[10pt,twocolumn]{article}

\usepackage[letterpaper, total={17.6cm,22.4cm}, columnsep=0.55cm]{geometry}
\usepackage[utf8]{inputenc}
\usepackage[T1]{fontenc}
\usepackage{amsmath, amssymb, bm, empheq}
\usepackage[T1]{tipa}
\usepackage{graphicx}
\usepackage[round]{natbib}
\usepackage{newtxtext, newtxmath}
\usepackage{fancyhdr}
\usepackage[font=small,labelfont=bf,justification=raggedright,
            singlelinecheck=false]{caption}
\usepackage[colorlinks=true, linkcolor=blue!50!black,
            citecolor=blue!50!black, urlcolor=blue!50!black]{hyperref}

\newcommand{\repobase}{https://github.com/FBerthommier/Vocalic-Spaces}
\newcommand{\repourl}{\href{\repobase}{\texttt{\repobase}}}

\newcommand{\linenomath}{}                 
\newcommand{\reprintcolumnwidth}{\columnwidth}
\newcommand{\email}[1]{}                   
\newcommand{\affiliation}[1]{}
\newcommand{\docsection}[1]{\section*{#1}}
\newcommand{\multimedia}[2]{\item[]\href{\repobase/multimedia/#1}{\texttt{#1}}
 ~-- #2\par}
\let\cite\citep                             

\fancypagestyle{plain}{%
  \fancyhf{}%
  \fancyhead[L]{\small\itshape A mathematical model of the vowel space}%
  \fancyhead[R]{\small Preprint --- arXiv:2111.00868v3}%
  \fancyfoot[C]{\small\thepage}%
  }
\begin{document}

\title{\bfseries A mathematical model of the vowel space}

\author{Fr\'ed\'eric Berthommier \\[2pt]
\small Univ. Grenoble Alpes, CNRS, Grenoble INP, GIPSA-lab, France
\thanks{Email: frederic.berthommier@gipsa-lab.grenoble-inp.fr.
Open-access preprint (two-column layout equivalent to the JASA-EL
submission), source: \href{https://arxiv.org/abs/2111.00868v3}
{arXiv:2111.00868v3}. Companion code and data: \repourl{}.
This edition is distributed under the MIT License.}}

\date{October 2021 (reformatted October 2026)}

\maketitle
\thispagestyle{plain}

\begin{abstract}
The articulatory-acoustic relationship is many-to-one and non-linear, which is a great limitation for the study of speech production. A simplification is proposed to establish a bijection between the vowel space $(f_1, f_2)$ and the parameter space of different vocal tract models. Following Schroeder's (1967) approach, the formant frequencies are related with the first two odd cosine Fourier coefficients of the area function. A generative model is proposed for building this relationship far from the neutral position and the use of the sensitivity function is avoided. Mixtures of cosines are realized allowing the generation of principal vowels: \textipa{[1, u, o, O, a, E, e, i]} with two simple formulas. This shows that these two latent variables bring together the geometry of the vocal tract and the vowel space. The mixing function is transformed into a coordination function $\bm{P}(\rho,\theta) = \bm{\Omega} + \rho\,\bm{\Psi_1}\, cos(\bm{\Psi_2} - \theta)$ that can integrate the geometric parameters of an articulatory model. Applied to the 4-tube DRM (Distinctive Region Model) derived from the generic model and to Fant's model, this function set a bijection between its input space and $(f_1, f_2)$ based on the two latent variables.
\end{abstract}

\section{\label{sec:1} Introduction}
Establishing a causal relationship between articulatory movements and speech sounds is a major goal of the speech sciences. The main theoretical advances were made during the 1960s, devoted to understanding the actual shapes of the vocal tract (VT) in sound production. This led to the design of articulatory models as well as to the synthesis of speech sounds, but this effort was slowed down by the complexity of this relationship. Even if we limit the domain to vowels and their first two formants $(f_1,f_2)$, this problem remains poorly understood. This is due to the many-to-one relation between the very large space of VT forms and the two-dimensional space of formants as well as to the non-linearities. Starting from real VT forms, this problem is so difficult that theoretical researches like \cite{Heinz1967,Atal1978,YehiaItakura1996} have failed to produce a clear conclusion. De facto, there is no reliable inversion technique applicable to any vocal tract, even with modern Bayesian or neural network approaches. We are therefore unable to clearly understand how speech evolved from the shapes of the monkey vocal tract, or to understand how movements are generated by the brain to control the vocal tract. This lack of understanding is at the root of long-standing conflicts such as the one examined and seemingly closed by \cite{Boe2019}. After the development of articulatory models such as Maeda's or Mermelstein's, the community split into two approaches: those who work directly with the area function to synthesize speech sounds and those who work with human-like models controlled by articulatory parameters. Both fields have thus seen many advances, but without much convergence. Until today, only the first approach has allowed to realize a bijection between the parametric space of the area function and the vowel space $(f_1, f_2)$ \cite{Carre2009,Story2005}. The aim of this paper is to revisit the roots in order to construct a link with articulatory modelling and to understand which causality and structure these models share. First, the technique adopted in section~\ref{sec:2} avoids real VT data in order to perfectly control the area function. This allows for a generation of vowel spaces that are formed from a seed of three well chosen vectors and not empirical functions derived from data as in \cite{Story2005}. In section~\ref{sec:3}, the simplest model, which we call the generic model, allows the synthesis of 8 vowels with two concise formulas having 2 variables and length as the only static parameter. Since the vowel space is filled from its periphery and not generated recurrently from its center as in \cite{Carre2004,Carre2009}, the use of the sensitivity function \cite{FantPauli1974,Carre2004,Story2006} is avoided. Thus, we free ourselves from any consideration of the validity of this principle and the generic model can acquire the status of an independent mathematical object. Moreover, the first generative function, which we call the three-phase mixing function, is transformed in section~\ref{sec:4} into a more comprehensible coordination function. This is the coordination of human-like articulators, which is tested in section~\ref{sec:5} with a 4-tube model controlled by articulatory parameters. The seed of this model is created according to intuitive principles. Comparison with the DRM \cite{Mrayati1988}, a well-known 4-tube model here derived from the generic model, provides insight into the role played by the coordination function in reducing the many-to-one relationship.

\section{\label{sec:2} The three-phase mixing function}
A generative function $\bm{P}(\theta)$ is constructed with the principle of the Gibbs triangle used in chemistry for the representation of ternary mixtures of components. The dimension of this graph is reduced to two because the three proportions $a$, $b$ and $c$ are reduced to a constant $K$ fixed at $1$ and the components are located at the three corners of the triangle. Since the vowel space is also considered a triangle with the extreme vowels \textipa{[u,a,i]} at the corners (see \cite{Boe2019}) and the neutral vowel \textipa{[@]} at the center, the generative function of the vocal tract shapes takes three vectors corresponding to these vowels as input and defines the neutral vowel as their average. While the acoustic space is a triangle, the vocal tract shape space (VT) will be defined as a circular domain with a center, this for each scalar value of a vector. Each vector of a cycle including the three vectors $\{\bm{i},\bm{j}, \bm{k}\}$ is given by the three-phase mixing function according to its angle $\theta$:
\linenomath
\begin{align}
\begin{split}
\bm{P}(\theta) = \frac {1}{3}\,(\bm{i}+\bm{j}+\bm{k}) + \frac {2}{3} \,(\bm{i}\,cos(\theta - \frac{\pi}{3}) + \\
\bm{j}\,cos(\theta - \pi) + \bm{k}\,cos(\theta - \frac{5\pi}{3})) \label{Eq1}
\end{split}
\end{align}
The first property of this function is to verify $\bm{P}(\frac{\pi}{3})=\bm{i}$, $\bm{P}(\pi)=\bm{j}$ and $\bm{P}(\frac{5\pi}{3})=\bm{k}$ and setting $\bm{\Omega}=\frac {1}{3}\,(\bm{i}+\bm{j}+\bm{k})$, the Eq.~\ref{Eq1} is rewritten:
\begin{subequations}
\begin{align}
\begin{split}
\bm{P}(\theta) = \bm{\Omega} + \frac {2}{3} \,(\bm{i}\,(cos\,\theta \,cos\,\frac{\pi}{3} + sin\,\theta \,sin\,\frac{\pi}{3}) \\
- \bm{j}\,cos\,\theta +\bm{k}\,(cos\,\theta \,cos\, \frac{5\pi}{3}+sin\,\theta \,sin\, \frac{5\pi}{3})),
\end{split}
\label{Eq2a} \\
\begin{split}
\rightarrow \bm{P}(\theta) = \bm{\Omega} + \frac {2}{3} \,(((\bm{i}+\bm{k})\,cos\,\frac{\pi}{3}\, -  \bm{j})\,cos\,\theta \\ +(\bm{i}-\bm{k}) \,sin\,\frac{\pi}{3} \,sin\,\theta).
\end{split}
\label{Eq2b}
\end{align}
\end{subequations}
This function is used as a generator of closed-open tube shapes, considering that $\{\bm{i},\bm{j}, \bm{k}\}$ are the vectors defining the shapes of the tubes producing \textipa{[u,a,i]} placed at given angles $\theta_h\in\{\frac{\pi}{3},\pi,\frac{5\pi}{3}\}$ of an equilateral triangle. The angle $\theta$ varies from $0$ to $2\pi$ in order to mix these components. Having a triangle on one side and a circle on the other is a main support for this paper. The aim is to address the nonlinear and multiple characteristics of the articulatory-acoustic relationship and to obtain a bijection between a large space $(f_1,f_2)$ of formant frequencies and a domain included in the very large VT shape space. This domain is constrained to incorporate the configurations of the 3 extreme vowels and to establish a continuum between them.

\section{\label{sec:3} The generic vocal tract model}

\begin{figure*}[ht]
\centering
\includegraphics[width=12cm,height=12cm]{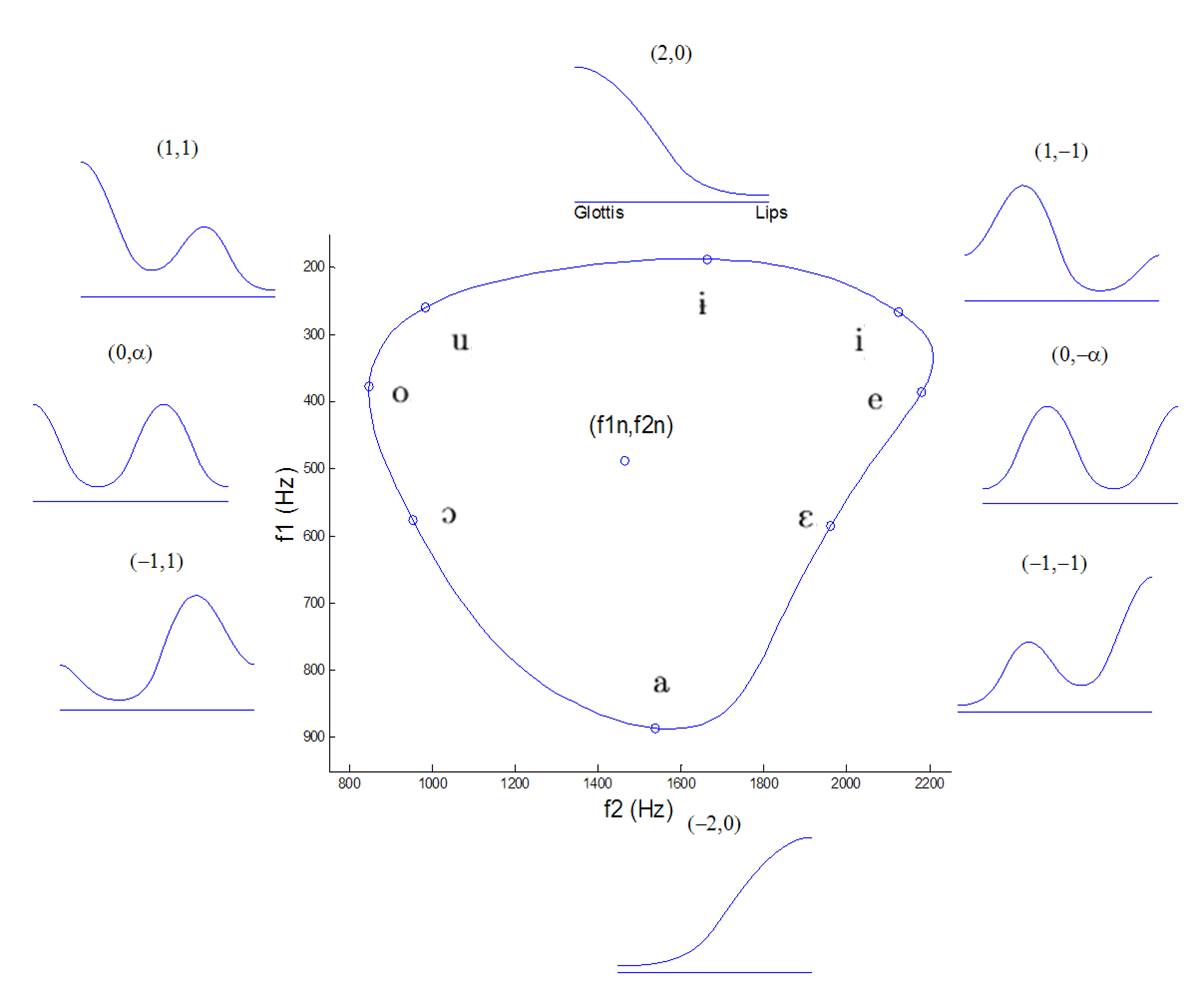}
\caption{\label{fig:FIG1}{The vowel space of the generic vocal tract model and its 8 characteristic vowels (Mm.~\ref{vowelsgeneric}). The area functions (Mm.~\ref{areageneric}) are plotted for each vowel with, at the top, the pairs $(a_1,a_2)$ of cosine Fourier coefficients which are introduced in the vowel equation (Eq.~\ref{Eq4a}b). For \textipa{[o,e]}, we have $\alpha=\tfrac {4}{3} \,sin\,\tfrac{\pi}{3}$.}}
\end{figure*}

Following the distinction proposed by \cite{StoryTitze1998} between VT modeling approaches, the objective here is to explore the formal counterpart of the empirical studies conducted by most researchers. The vectors $\{\bm{i},\bm{j}, \bm{k}\}$ are chosen from a theoretical point of view and not in close relation to real forms of VT. The first objective is to establish a link between the physics of closed-open tubes and the structure of the vowel space, and then to propose a unification of the two approaches. \cite{Schroeder1967} established from the so called Ehrenfest's theorem $\tfrac {\delta f_i}{f_i} = \tfrac{\delta E_i}{E_i}$ that for small perturbations of the neutral tube, the first 2 odd cosine Fourier coefficients of the area function are related to the resonant frequency variations: $\tfrac {\delta f_i}{f_i} = -\tfrac {1}{2} a_i$ ($a_1$ being the amplitude of $cos(\tfrac {\pi x}{L})$ and $a_2$ of $cos(\tfrac {3 \pi x}{L})$). This has a physical meaning independent of a specific transfer function between the VT shape space and the acoustic space, and others \cite{Mermelstein1967,Heinz1967} have derived the same relation from Webster's Horn equation after \cite{Ungeheuer62}. Let us evaluate these terms in the three-phase mixtures provided by Eq.~\ref{Eq1}. To simplify the mathematical expressions, $cos(\tfrac {\pi x}{L})$ is renamed to $\bm{v_1}$ and $cos(\tfrac {3 \pi x}{L})$ to $\bm{v_2}$. Then $\{\bm{i},\bm{j}, \bm{k}\}$ are defined as $\{\bm{1}+\bm{v_1}+\bm{v_2},\bm{1}-2\,\bm{v_1}, \bm{1}+\bm{v_1}-\bm{v_2}\}$. These are the simplest components compatible with the three-phase mixing function Eq.~\ref{Eq1} and at the same time able to maximize and minimize the first two frequency resonances by extrapolation of the Schroeder-Ehrenfest relation. First, this function satisfies the condition $\bm{\Omega}=\tfrac{1}{3}(\bm{i}+\bm{j}+\bm{k})=\bm{1}$. Then, these terms are introduced in Eq.~\ref{Eq2b} to identify $a_1(\theta)$ and $a_2(\theta)$:
\linenomath
\begin{subequations}
\begin{align}
&\rightarrow \bm{P}(\theta) = \bm{1} + 2\, \bm{v_1} \,cos\,\theta + \frac {4}{3} \, \bm{v_2}  \,sin\,\frac{\pi}{3} \,sin\,\theta,  \label{Eq3a} \\
\begin{split}
&\text{then} \, a_1(\theta)=2\,cos\,\theta, \, a_2(\theta)=\frac {4}{3} \,sin\,\frac{\pi}{3} \,sin\,\theta \, \\
&\text{with} \, \bm{P}(\theta) = \bm{1} + a_1(\theta) \, \bm{v_1} + a_2(\theta) \, \bm{v_2}.\label{Eq3b}
\end{split}
\end{align}
\end{subequations}
For main vowel synthesis, there is no other parameter than the length $L$ given in $cm$ and the amplitude is already scaled in $cm^2$. A suitable soft rectifier Eq.~\ref{Eq4b} is applied to the continuous functions $P(x,\theta)$ arising from Eq.~\ref{Eq3b} to obtain positive valued area functions $A(x,\theta)$. This is the only explicit nonlinearity introduced in the model that can distort the Schroeder-Ehrenfest relation away from the neutral tube. These are sampled to apply the transfer function of a transmission line model. For the present simulations\footnote{Provided in the companion repository: Matlab supplement \mbox{simuGEN.m} and validated Python translation.}, we chose the Badin and Fant model \cite{BadinFant1984} defined as lossless. The choice of transmission line model is not critical to reproduce the present results with these equations. For the sampling rate, a large value of $n\ge 100$ tubelets of length $\tfrac{L}{n}$ is preferable to avoid the distortions introduced at this stage. The length $L$ is fixed at $17.5\, cm$ for comparison with other works. The resulting concise vowel equation is as follows:
\linenomath
\begin{subequations}
\begin{empheq}[left=\empheqlbrace]{alignat=2}
\begin{split}
A(x,\theta) = expif(1 + a_1(\theta)\,cos(\frac {\pi x}{L}) + \\ a_2(\theta)\,cos(\frac {3 \pi x}{L})),
\end{split}
\label{Eq4a} \\
\begin{split}
\text{if} \, (y < 1) \, \text{then} \, expif(y)=exp(y-1) \, \\
\text{else} \, expif(y)=y.
\end{split}
\label{Eq4b}
\end{empheq}
\end{subequations}
This equation associates the main vowels with $\theta$ angles corresponding to pairs of Fourier coefficients $a_1$ and $a_2$. In Fig.~\ref{fig:FIG1}, the series of eight vowels \textipa{[1, u, o, O, a, E, e, i]} is produced with $\theta\in\{0,\tfrac{\pi}{3},\tfrac{\pi}{2},\tfrac{2\pi}{3},\pi,\tfrac{4\pi}{3},\tfrac{3\pi}{4},\tfrac{5\pi}{3}\}$ and with Fourier coefficients (partially) given by Eq.~\ref{Eq3b}  $(a_1,a_2)\in\{(2,0),(1,1),(0,\alpha),(-1,1),(-2,0),
(-1,-1),(0,-\alpha),(1,-1)\}$ with $\alpha=\tfrac {4}{3} \,sin\,\tfrac{\pi}{3}$. It is logical to find for \textipa{[u, a, i]} the coefficients given for the definition of $\{\bm{i},\bm{j}, \bm{k}\}$. Let us mention the analogy with an octatonic musical scale having W-H-H-W-W-H-H-W intervals which suggests that a common structure to speech and music is revealed by this representation. Moreover, \cite{FenkOczlon2017} discussed \textit{correspondences between the number of vowels and the number of pitches in musical scales across cultures} and we add that vowels might have a compatible scale. Equation.~\ref{Eq1} has a second important property for the expression of the structure of the vowel space which is the antisymmetry with respect to $\bm{\Omega}$: we have $\bm{P}(\theta+\pi) = 2\,\bm{\Omega} - \bm{P}(\theta)$ then the coefficients of \textipa{[E, 1, O]} can be directly deduced from those of \textipa{[u, a, i]} because $a_i(\theta+\pi)=-a_i(\theta)$. The extrema of the acoustic space are well reached in spite of the nonlinearities transforming the circle into a triangular shape (Fig.~\ref{fig:FIG1}): \textipa{[1, a, o, e]} correspond well to $\{min(f_1),max(f_1),min(f_2),max(f_2)\}$ and this is determined by the values of $a_i$. To generate the complete vowel space, the extension to a domain is performed by multiplying $a_1(\theta)$ and $a_2(\theta)$ by a scalar $\rho$ varying between $0$ and $1$. Thus, the central vowel \textipa{[@]} corresponds to $\rho=0$ and except for this point, a bijection is established between the domains $\bm{P}(\rho,\theta) \leftrightarrow (\rho\,a1,\rho\,a2)$ and this is observed between all of them and the vowel space $(f_1,f_2)$. The variation of $(f_1,f_2)$ are consistent with those of $(\rho\,a1,\rho\,a2)$ and these become the key variables which are here directly controlled.

\section{\label{sec:4} The coordination function}
\begin{figure}[ht]
\centering
\includegraphics[width=\reprintcolumnwidth]{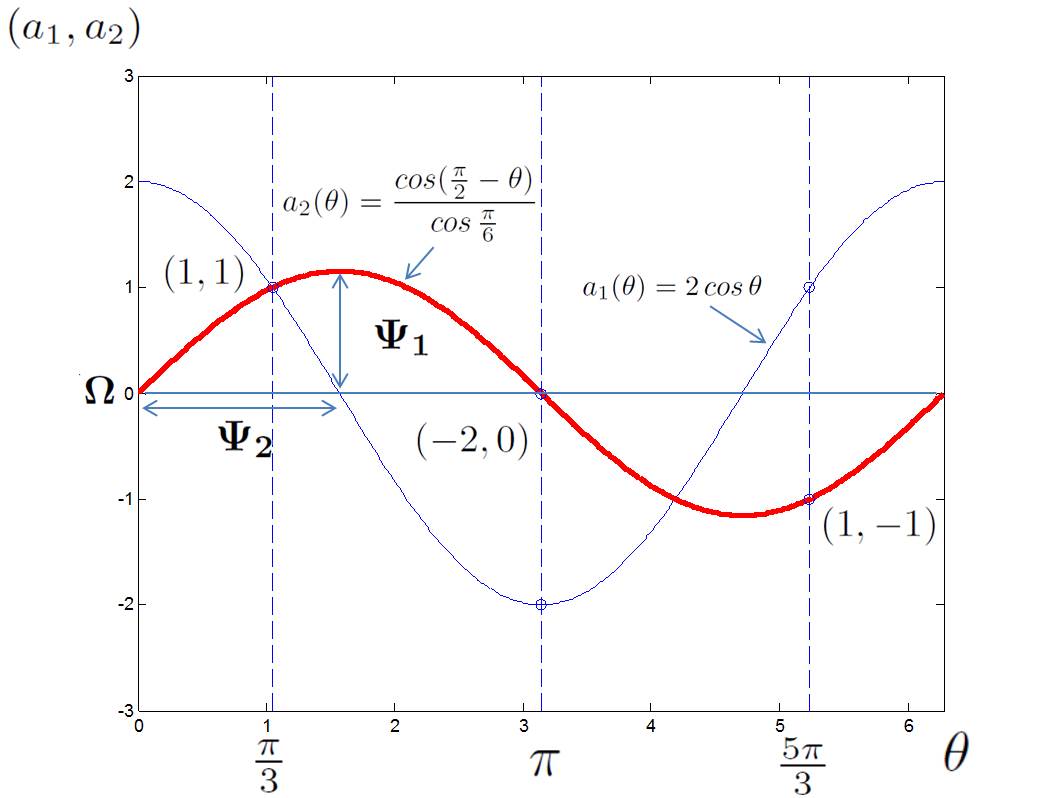}
\caption{\label{fig:FIG2}{Plot of the functions $(a_1,a_2)$ expressed with the values of $\{\bm{\Omega},\bm{\Psi_1},\bm{\Psi_2}\}$ derived in Table~\ref{tab:tab1}. This shows how $\{\bm{i},\bm{j}, \bm{k}\}$ are set at their respective phases and it focuses on $a_2(\theta)$ for $\{\bm{\Omega},\bm{\Psi_1},\bm{\Psi_2}\}$. The expression of $a_2(\theta)$ is consistent with this given by Eq.~\ref{Eq3b} because $cos(\frac{\pi}{2} - \theta)=sin \, \theta $ and $cos^{-1} \tfrac{\pi}{6}=\alpha$.}}
\end{figure}

Another more practical equation for n-tubes and articulatory modeling can be rewritten from the three-phase mixing function (Eq.~\ref{Eq1}). Let us insist that this is mathematically equivalent. This highlights the properties of the resulting parameter vector $\bm{P}(\theta)$ and its terms are significant: they express the coordination between the parameters along the $\theta$ cycle. In other words, for articulatory modeling, they explain how the phases and amplitudes of the different parameters of a model are locked together in a continuous domain. The area function is a side product that is analyzed section~\ref{sec:5} to find a link between these two classes of models. As developed in section~\ref{sec:5}, it is also possible to apply it not only with the vectors $\{\bm{i},\bm{j}, \bm{k}\}$, but also with their three dual vectors $\{\bm{\Omega},\bm{\Psi_1},\bm{\Psi_2}\}$ appearing in this function. As it is a rewriting, we give a priori the new form in which the two unknown terms are identified from those of Eq.~\ref{Eq2b} with element wise operations (we already know the third $\bm{\Omega}=\tfrac {1}{3}\,(\bm{i}+\bm{j}+\bm{k})$):
\linenomath
\begin{subequations}
\begin{align}
&\bm{P}(\theta) = \bm{\Omega} + \bm{\Psi_1}\, cos(\bm{\Psi_2} - \theta) \label{eq5a} \\
&\rightarrow \, \bm{P}(\theta) = \bm{\Omega} + \bm{\Psi_1}\,(cos\,\bm{\Psi_2} \,cos\,\theta + sin\,\bm{\Psi_2} \,sin\,\theta) \label{eq5b}
\end{align}
\end{subequations}
After identification, we obtain:
$\bm{\Psi_1}\, cos\,\bm{\Psi_2}=\tfrac {2}{3} \,((\bm{i}+\bm{k})\,cos\,\tfrac{\pi}{3}\, -  \bm{j})$ and $\bm{\Psi_1}\, sin\,\bm{\Psi_2}=\tfrac {2}{3} \,(\bm{i}-\bm{k}) \,sin\,\tfrac{\pi}{3}$. At first, the values of the vector $\bm{\Psi_2}$ are derived from those of $\{\bm{i},\bm{j}, \bm{k}\}$:
\linenomath
\begin{equation}
\bm{\Psi_2}=atan(((\bm{i}-\bm{k}) \,sin\,\frac{\pi}{3})/(\frac{1}{2}\,(\bm{i}+\bm{k})-\bm{j}))\label{Eq6}
\end{equation}
Then, we remark that $\bm{\Psi_1} =(\bm{h}-\bm{\Omega})\,cos^{-1}(\bm{\Psi_2}-\theta_h) \,\text{with} \, \bm{h}\in\{\bm{i},\bm{j}, \bm{k}\}$ but this expression is not always valid for the same $\bm{h}$ for the whole vector. Starting with a given $\bm{h}$, this is necessary to check element by element if $\bm{h}_i = \bm{\Omega}_i$ and to select another one when it is true in order to find a non zero solution. To illustrate this procedure, we show how to set the coordination function for $(a_1,a_2)$ themselves and the result is in Table~\ref{tab:tab1}. The values of $\{\bm{i},\bm{j}, \bm{k}\}$ are in Fig.~\ref{fig:FIG1}. The calculation of $\bm{\Psi_2}$ is achieved with Eq.~\ref{Eq6} and the checking is applied for the two elements sequentially. Starting with the default $\bm{h} = \bm{j}$, the test is false for $a_1$ and then $\bm{\Psi_1}$ acquires the value $(-2 - 0)\,cos^{-1}(0-\pi)=2$. For $a_2$, the test is true and we have to change for $\bm{h} = \bm{i}$ (or $\bm{h} = \bm{k}$) in order to find $(1 - 0)\,cos^{-1}(\frac{\pi}{3}-\frac{\pi}{2})=cos^{-1}\frac{\pi}{6}$. The resulting functions are plotted Fig.~\ref{fig:FIG2} and this shows that each the two cosines passes through the three original points placed at given angles $\theta_h\in\{\frac{\pi}{3},\pi,\frac{5\pi}{3}\}$. Then, the two parameters are embedded in the same cycle. The goal of the transformation is to get this cycle in a space of parameters describing the shape of the vocal tract whatever the type of this description. This can also be the parameters of a Maeda's model or more generally every parameters varying the geometry of the vocal tract.

The antisymmetry with respect to $\bm{\Omega}$ is easily proved in this second form: we have $\bm{P}(\theta+\pi) = 2\,\bm{\Omega} - \bm{P}(\theta)$ since $cos(\bm{\Psi_2} - (\theta+\pi))=-cos(\bm{\Psi_2} -\theta)$. In particular, the effect of each scalar parameter included in $\bm{P}$ at each $\theta$ is easy to identify as agonist or antagonist and this is useful for controlling articulatory models. The variations of each parameter along the cycle can be coded according to a rule for assigning a parameter extremum to a phase. For example, the length variation is coded as $L=\overline{L}+\Delta L\,cos(\tfrac{\pi}{3}-\theta)$ for having a maximum $\overline{L}+\Delta L$ for the phase $\tfrac{\pi}{3}$ of vowel \textipa{[u]}. Thus, the direct method is the conversion of the known vectors of the parameters of \textipa{[u,a,i]} with Eq.~\ref{Eq6} and the previous procedure. As before, the coordination function is extended to a domain by multiplying the amplitude $\bm{\Psi_1}$ by $\rho$:
\linenomath
\begin{equation}
\bm{P}(\rho,\theta) = \bm{\Omega} + \rho\,\bm{\Psi_1}\, cos(\bm{\Psi_2} - \theta) \label{Eq7}
\end{equation}
This formula is easy to use and can be applied to real VT data such as those published by \cite{Story2005}. From 3 extreme vowel vectors including the length component, it generates a mapping comparable to that obtained with empirical orthogonal modes but in a much simpler way. Nevertheless, the control of the area function is not the goal of the coordination function because it is already fulfilled by the three-phase mixing function. Applied to the area function, the new parameters of the coordination function have little meaning but, on the contrary, they express well the variations of geometrical parameters such as the length of the VT. Polar coordinates are particularly well suited to formalize the movements of the tongue. The next step is to show that the integration of the geometric variations in the coordination function leads to the same variations of $(a_1,a_2)$ as well as to the same relations between $(a_1,a_2)$ and $(f_1,f_2)$. In what follows, the cosine Fourier coefficients are not generated directly but observed as latent variables $(\tilde{a}_1, \tilde{a}_2)$.

\section{\label{sec:5} Comparison of two 4-tube models}
\begin{figure*}[ht]
\centering
\includegraphics[width=11.7cm,height=11.7cm]{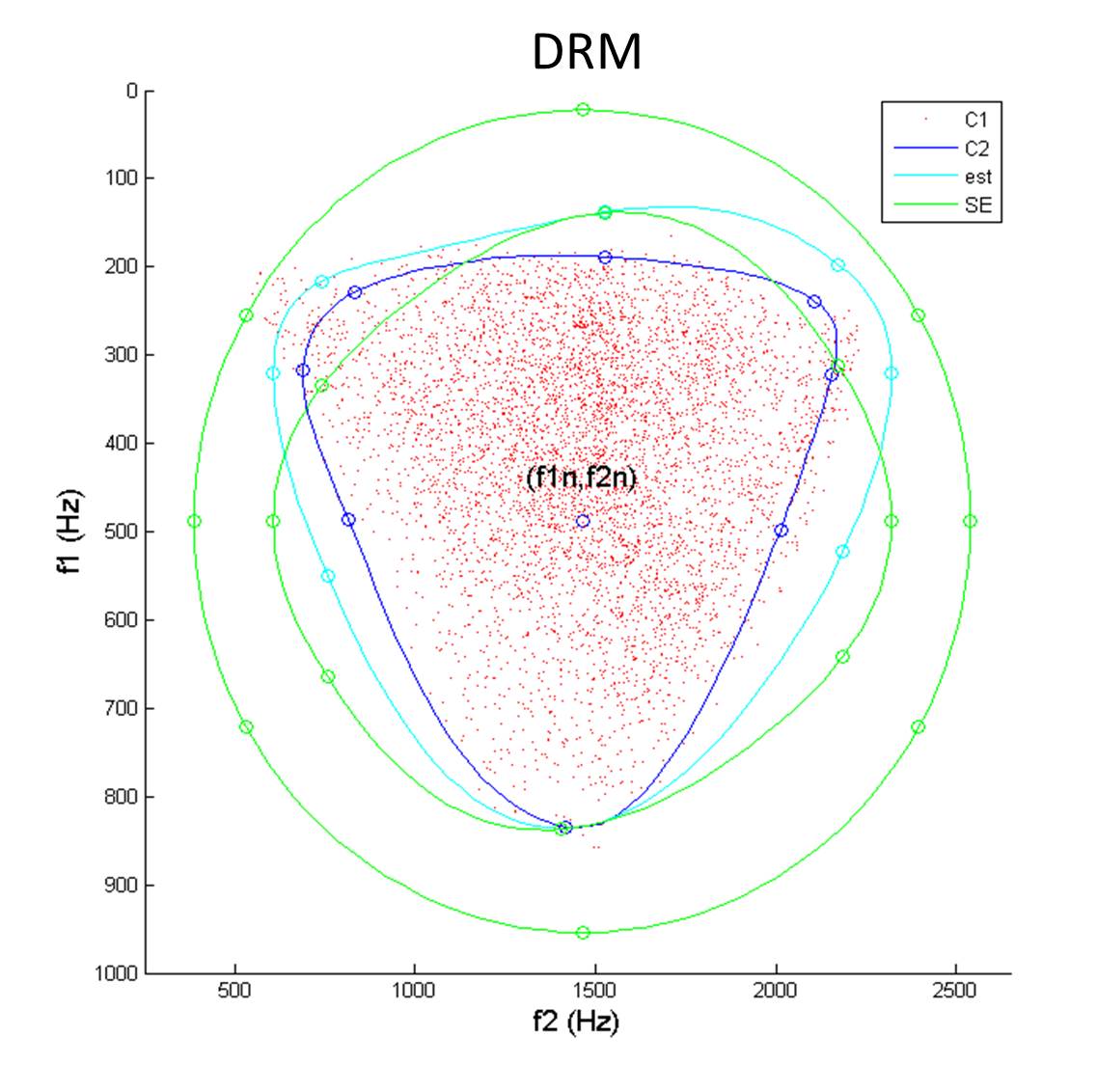}
\caption{\label{fig:FIG3}{Simulation in condition $\bm{C1}$ together with the vowel space $\bm{C2}$ with $\rho=1$ obtained with the coordination function (Mm.~\ref{areaDRM} and Mm.~\ref{vowelsDRM}). All (non figured) points of $\bm{C2}$ are inside of this. The spaces derived from the Schroeder-Ehrenfest relation and Fourier coefficients before and after rectification ("SE" from $(a_1,a_2)$ and $(\tilde{a}_1, \tilde{a}_2)$) and the biased relation ("est" from  $(\tilde{a}_1, \tilde{a}_2)$) are plotted relative to the neutral reference $(f_{1n},f_{2n})$. The bias transforms the circular shape "SE" from $(\tilde{a}_1, \tilde{a}_2)$ into the triangular shape "est".}}
\end{figure*}

\begin{figure*}[ht]
\centering
\includegraphics[width=16cm,height=8cm]{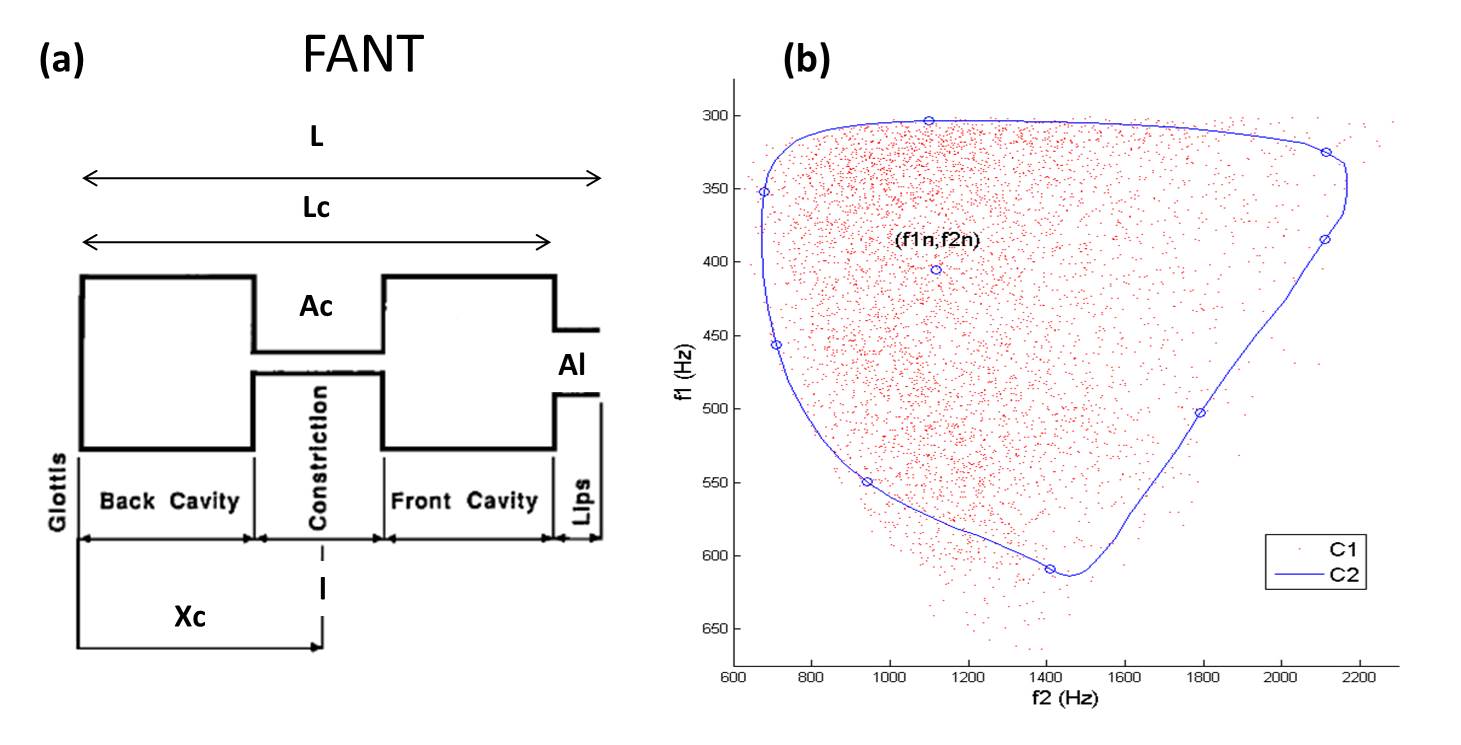}  
\caption{\label{fig:FIG4}{(a) The design of the Fant model redrawn from \cite[Fig.~1]{Badin1990} and the definition of geometrical parameters used in Table~\ref{tab:tab3} (b) Simulation in condition $\bm{C1}$ together with the vowel space $\bm{C2}$ with $\rho=1$ obtained with the coordination function (Mm.~\ref{areaFant} and Mm.~\ref{vowelsFant}). All (non figured) points of $\bm{C2}$ are inside of this.}}
\end{figure*}

While the generic model has 2 degrees of freedom to generate a large vowel space $(f_1,f_2)$, the 4-tube models \cite{StevensHouse1955,Fant1960} require 3 homogeneous parameters with $\{X_c,A_c,A_l\}$ (constriction position and area, lips area). It has been shown that anthropomorphic models also require at least these parameters for inversion \cite{Boe1992}. On the other hand, with Monte Carlo simulations with 4-tube with a total length set at $17.5\, cm$, \cite{Boe2019} noticed that the vowel space is maximized with 7 parameters and that having $N>4$ sections is unnecessary. This maximized vowel space was defined as the articulatory potential of a closed-open tube having a given length. The generic model has somewhat smaller $(f_1,f_2)$ (Fig.~\ref{fig:FIG1}) and we suspect that the extra $5$ of parameters are the source of the undesirable many to one relationship, partly due to the fact that the 4-tube cut is determined by 3 of them. Note that the third formant is not incorporated into the current definition of vowel space. Therefore, to obtain its control, the generic model must be defined with $cos(\frac {5 \pi x}{L})$ terms leading to an additive parameter. This would be consistent with $\{X_c,A_c,A_l\}$ but this improvement is to be developed later. We derive from the generic model the well-known 4-tube DRM having an optimal cut \cite{Mrayati1988} and compare it with a circularized Fant model with 4 parameters $\{X_c,A_c,A_l,L\}$.

The goal of the DRM is to propose a division into 8 regions and a rule to vary their areas by knowing the direction of variation of each of the first 3 formants. For the 2 formants only, we limit ourselves to 4 regions and the cut of the closed-open tube is done at $\{\tfrac{L}{6},\tfrac{L}{2},\tfrac{5 L}{6}\}$ from glottis to lips. The parameters of the DRM are the areas of each section named $\{A_1,A_2,A_3,A_4\}$ while the corresponding values of the generic model are the $P_i$ values before rectification (Eq.~\ref{Eq4b}). Since the cutting is done at zero crossings of the $\bm{v_1}$ and $\bm{v_2}$ components, these values are taken in the generic model at the $x\in\{0,\tfrac{L}{3},\tfrac{2 L}{3},L\}$ points to compute $P_i$. Moreover, the values of $\{P_1,P_2\}$ are antisymmetric by construction from the generic model with those of $\{P_3,P_4\}$. This is an internal antisymmetry about the midpoint $\tfrac{L}{2}$, well described by \cite{Mrayati1988} and not to be confused with that created by the coordination function, although it has common effects. Because $v_i(0)=-v_i(L)$ and $v_i(\tfrac{L}{3})=-v_i(L-\tfrac{L}{3})=-v_i(\tfrac{2 L}{3})$ we obtain $P_3=2-P_2$ and $P_4=2-P_1$ from Eq.~\ref{Eq3b}. What is called the reduction is the systematic calculation of $\{P_3,P_4\}$ from $\{P_1,P_2\}$ so that this property explicitly reduces the number of free parameters to 2. When this reduction is not applied, the coordination function preserves the internal antisymmetry introduced by the components $\bm{v_1}$ and $\bm{v_2}$. This is because their amplitudes respect $a_i(\theta+\pi)=-a_i(\theta)$ along the cycle as seen in section~\ref{sec:3}. There is an equivalence between the explicit reduction and the coordination and this is encoded in $\bm{\Psi_1}$ and $\bm{\Psi_2}$. We find here the main property of the coordination function which consists in relying on the odd cosine Fourier components. With the generic model and the present DRM reduction, these 2 components are controlled variables so that only two degrees of freedom are needed to determine the variations of $(f_1,f_2)$ away from the neutral tube, even if they are non linear. This is a baseline for observing the effect of the coordination function when the $\bm{v_1}$ and $\bm{v_2}$ components and many others are not controlled. Its configuration\footnote{Provided in the companion repository: Matlab supplement \mbox{simuDRM.m} and validated Python translation.} for $\{P_1,P_2\}$ after the transformation $\{\bm{i},\bm{j}, \bm{k}\}\rightarrow\{\bm{\Omega},\bm{\Psi_1},\bm{\Psi_2}\}$ is detailed in Table~\ref{tab:tab2}. Following the same notation as for the generic model, for $\{P_3,P_4\}$, we have $\Psi_1(\tfrac{2 L}{3})=-\Psi_1(\tfrac{L}{3}),\,\Psi_2(\tfrac{2 L}{3})=\Psi_2(\tfrac{L}{3})$ and  $\Psi_1(L)=-\Psi_1(0),\,\Psi_2(L)=\Psi_2(0)$. This is easy to check that $P_3=2-P_2$ and $P_4=2-P_1$ as for the reduction.

The version of Fant's model that was implemented\footnote{Provided in the companion repository: Matlab supplement \mbox{simuFANT.m} and validated Python translation.} by \cite{Badin1990} is embedded in the coordination function format. Its main difference is that a constant length section is sliding (Fig.~\ref{fig:FIG4}a) whereas the DRM is a piston model with a fixed cut: it varies the areas of sections only. The advantage pointed out by \cite{BoePerrier1990} is to represent the tongue movements by variations of $\{X_c,A_c\}$ and this has been considered until now as a divergence between these two classes of models. However, the geometry of this model does not have the symmetries introduced by the cut and at the same time it reduces the antisymmetry of the area function. This does not mean that the latter is unrecoverable or that it has no role. Indeed, the vocal tract does not function as a piston model but there is no evidence that they have nothing in common as defended by \cite{BoePerrier1990}. To equalize their number of parameters to 4, we also vary the length $L$ in the Fant model. We have seen in section~\ref{sec:4} how to find $\{\bm{\Omega},\bm{\Psi_1},\bm{\Psi_2}\}$ for this parameter. The principle is the same for the 3 others: (1) choice of the offset (2) fixing the variation amplitude around the offset before soft rectification (3) assigning an extremum of each parameter to a vowel phase. For example Table~\ref{tab:tab3}, $X_c$ is maximal for \textipa{[i]} so that the configuration acquires a large posterior Helmholtz cavity specific to this vowel \cite{Boe2019}. Note that the language adopted to specify configurations is different in articulatory modeling because it is goal-oriented and qualitative. Based on the geometry of the model, the coordination function specifies how the parameters are cycled. Here, variations in $\{X_c,A_c\}$ correspond to an anatomically relevant rotation of the tongue in the mouth. This is also easy to recover $\{\bm{i},\bm{j}, \bm{k}\}$ from $\{\bm{\Omega},\bm{\Psi_1},\bm{\Psi_2}\}$. At this point, the coordination function is fully compatible with both classes of models. Nevertheless, there is a reconstruction step to have a sampled $P(i,\rho,\theta)$ which is described in the captions of Table~\ref{tab:tab2} and \ref{tab:tab3}. Then, soft rectification is applied with Eq.~\ref{Eq4b}, changing the DCT spectrum for both models. Remarkably, a small amount of even components are introduced into the area function of the DRM. The transmission line model is set lossless for the DRM and lossy for the Fant model.

\begingroup
\setlength{\abovedisplayskip}{3pt plus 1pt minus 1pt}%
\setlength{\belowdisplayskip}{3pt plus 1pt minus 1pt}%
The models are first simulated at $\rho=1$ and $\theta$ varying from $0$ to $2\pi$ with the coordination function (labelled $\bm{C2}$ in Fig.~\ref{fig:FIG3} and \ref{fig:FIG4}). Both vowel spaces have a similar structure with the 8 vowels placed comparatively (Fig.~\ref{fig:FIG3}, \ref{fig:FIG4}b). The size of the vowel space for Fant's model appears to be reduced. The two 4-tube models are simulated under 2 conditions: (1) $\bm{C1}$, with their 4 parameters uncorrelated and uniformly distributed within $\bm{\Omega}\pm\bm{\Psi_1}$ (2) $\bm{C2}$ using the coordination function with $\rho=1$ together with $(\rho,\theta)$ randomnesses. We observe Fig.~\ref{fig:FIG3} that for the DRM, the space $(f_1,f_2)$ is identically covered except at the corners of the uniform distribution. For Fant's model, more points of $\bm{C1}$ escape from the $\bm{C2}$ space but this is globally confined. Then $\bm{C2}$ does not decrease so much the articulatory potential of Fant's model as defined by $\bm{C1}$. We evaluate the two cosine Fourier coefficients of the area function which are here the most informative components of the DCT spectrum:
\linenomath
\begin{subequations}
\begin{empheq}[left=\empheqlbrace]{alignat=2}
&\tilde{a}_1(\rho,\theta)=\frac{2}{n}\sum_{i=1}^n A(i,\rho,\theta)\,cos(\frac {\pi (i-\frac{1}{2})}{n}), \label{eq8a} \\
&\tilde{a}_2(\rho,\theta)=\frac{2}{n}\sum_{i=1}^n A(i,\rho,\theta)\,cos(\frac {3\pi (i-\frac{1}{2})}{n}). \label{eq8b}
\end{empheq}
\end{subequations}%
\endgroup

\begin{figure*}[ht]
\centering
\includegraphics[width=16cm,height=8cm]{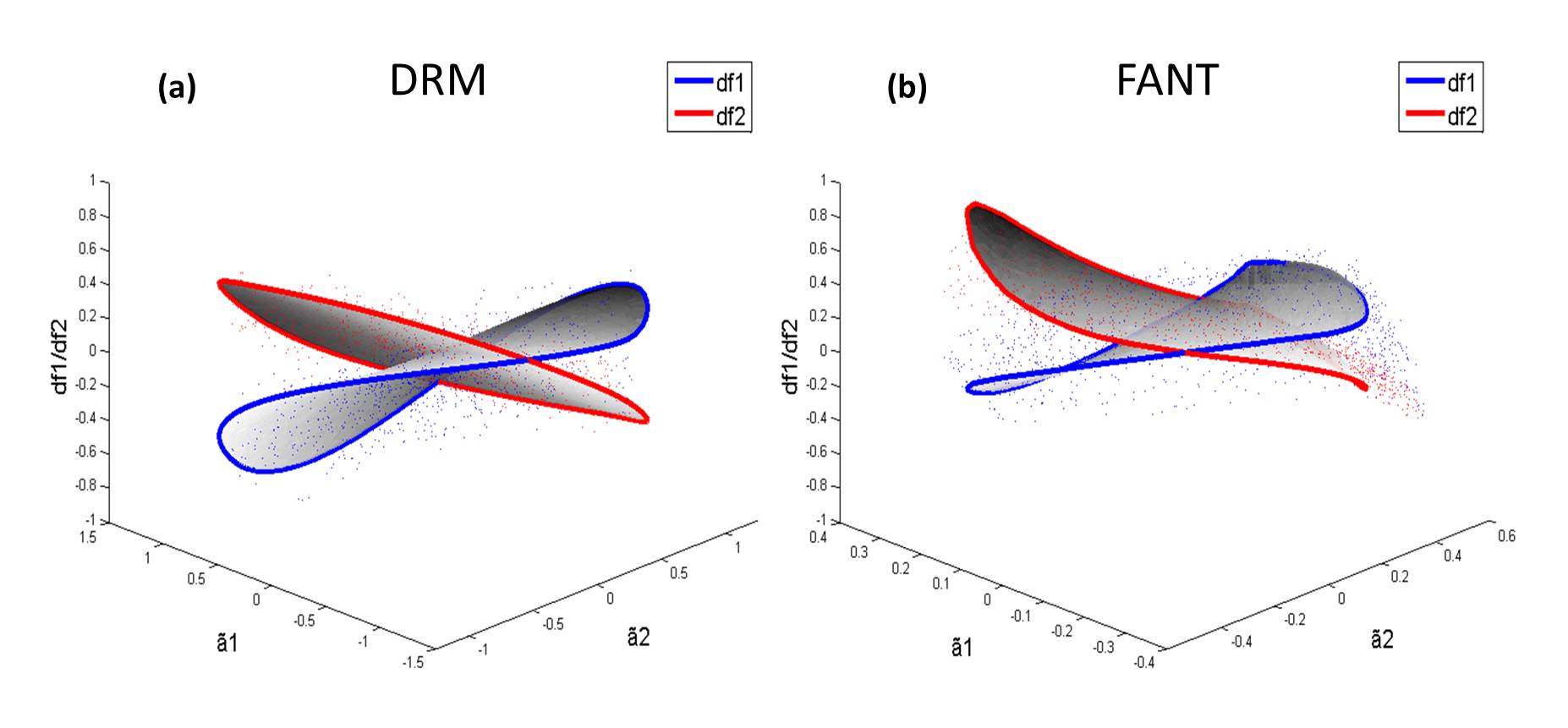}
\caption{\label{fig:FIG5}{Simulations of $df_i$ in condition $\bm{C1}$ and $\bm{C2}$: (a) DRM (b) Fant model. Abscissa: Fourier cosine coefficients of the area function. Ordinate: Deviations from neutral in relative values $df_i=(f_i-f_{in})/f_{in}$ for the two first formants. The circled gray surfaces represent the configurations produced by the coordination function in $\bm{C2}$ and the dots represent $\bm{C1}$ which is a simulation of the 4 parameters of each model but uncorrelated in the same domain of variation $\bm{\Omega}\pm\bm{\Psi_1}$.}}
\end{figure*}

For each of the models, the formants $(f_{1n},f_{2n})$ of the neutral position $\bm{P}(0,0)$ are calculated. Let remark this is not an uniform area function for the Fant model. In Fig.~\ref{fig:FIG5}, the deviations from the neutral position in relative values $df_1=(f_1-f_{1n})/f_1$ and $df_2=(f_2-f_{2n})/f_2$ are plotted as a function of $(\tilde{a}_1,\tilde{a}_2)$. This shows that $\bm{C2}$ is represented by surfaces whereas $\bm{C1}$ is represented by disparate dots distant from these surfaces. Considering that $(\tilde{a}_1,\tilde{a}_2)$ reflects control parameters, the key point is that we observe for $\bm{C1}$ the many to one relationship whereas it disappears for $\bm{C2}$: the intrinsic dimension is reduced at 2 for both models. By combining the two surfaces which cross in each graph, a bijection is established between pairs $(\tilde{a}_1,\tilde{a}_2)$ and $(df_1,df_2)$. Note, however, that there is a non-visible deviation from bijectivity between \textipa{[a]} and \textipa{[O]} for Fant's model, but that this does not affect the validity of this observation. For DRM, we have seen that the dimension can be explicitly reduced due to internal antisymmetry. Here, the only reason to have a many-to-one relationship is the decorrelation between the areas $A_i$. The coordination function only generates the internal antisymmetry as does the explicit reduction. Moreover, the variations of $(f_1, f_2)$ according to $(\tilde{a}_1,\tilde{a}_2)$ from one model to another are well correlated in the $\bm{C2}$ condition. For the Fant model, the coordination function generates a geometrically consistent set of configurations that are antisymmetric because the latent variables $(\tilde{a}_1,\tilde{a}_2)$ vary circularly. Despite the weaker internal antisymmetry and then weaker values of $(\tilde{a}_1,\tilde{a}_2)$, the two vowel spaces acquire a similar structure. The first explanation for the disparity of the $\bm{C1}$ points in Fig.~\ref{fig:FIG5}a, which is the signature of the many-to-one relation, is the growth of the even cosine Fourier coefficients associated with a degree of internal symmetry. \cite{Schroeder1967} showed as a corollary that these coefficients have no first order effect on $(f_1,f_2)$ and we extrapolate here by assuming that they have a small effect in the range of odd coefficients. After \cite{Mermelstein1967}, \cite{YehiaItakura1996} tried to disentangle them using constraints, knowing they cannot be evaluated from the formants $f_i$. Then, the decorrelation between areas $A_i$ in DRM $\bm{C1}$ is a source of many-to-one relationship as shown in Fig.~\ref{fig:FIG5}a.
\begin{figure}[ht]
\centering
\includegraphics[width=\reprintcolumnwidth]{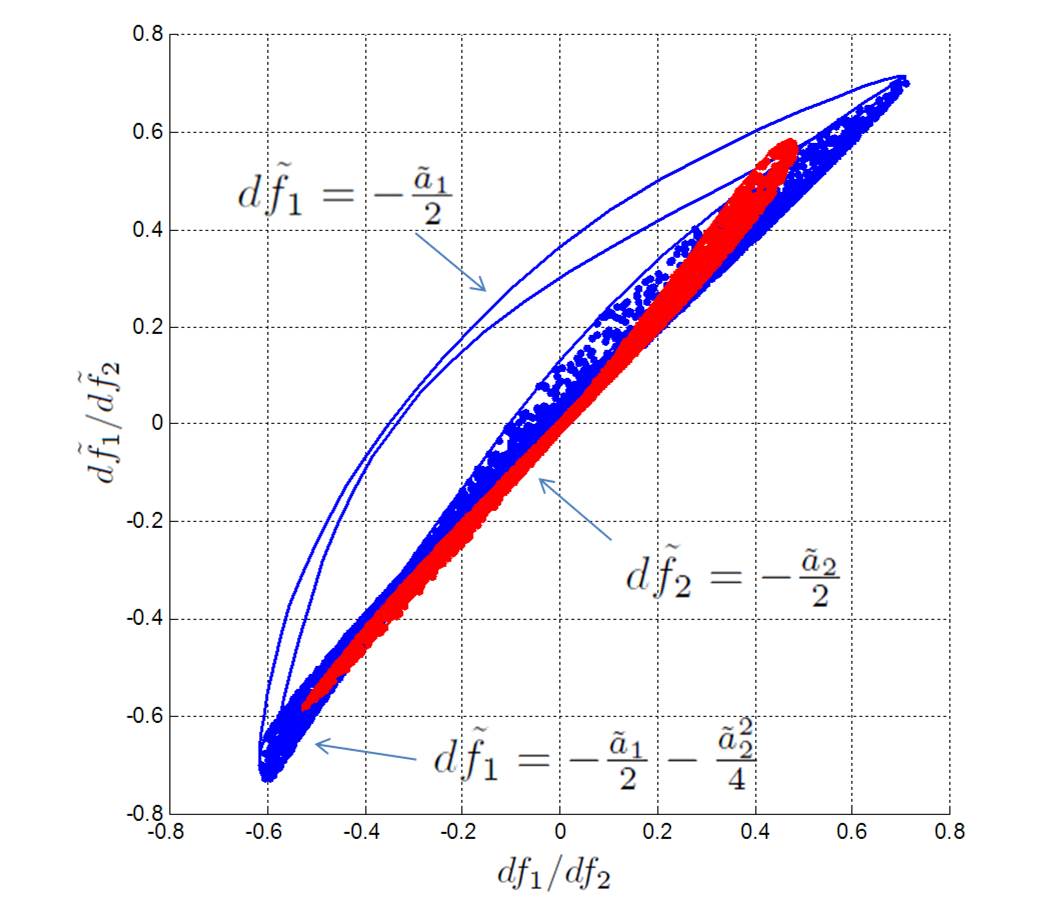}
\caption{\label{fig:FIG6}{Plot of the estimates of the deviations from neutral in relative values for DRM $\bm{C2}$. For the first formant, this is close to the bisecting line thanks to the bias.}}
\end{figure}

A remaining question is the relationship between the resonant frequencies and the cosine Fourier coefficients. It turns out that for the generic model as well as for the reduced DRM, this presents a quadratic term biasing the Schroeder-Ehrenfest relation only for $df_1$. The variations away from the neutral tube are approximated fairly well by $d\tilde{f}_1=-\tfrac{\tilde{a}_1}{2}-\tfrac{\tilde{a}_2^2}{4}$ and $d\tilde{f}_2=-\tfrac{\tilde{a}_2}{2}$ (Fig.~\ref{fig:FIG6}). As shown in Fig.~\ref{fig:FIG3}, this has the main effect of transforming a circular shape into a triangular shape, providing physical support for the common shape of the vowel space. This cannot be due to soft rectification because it does not produce such a bias but it is an interference in the transfer function. This fitting is a surprising simplification to explain. In \cite[Fig.~3 and Fig.~4]{Carre2004}, the effect of independent variations of $\tilde{a}_1$ and $\tilde{a}_2$ is well described. The goal was to reach the extremes of the vowel space using an iterative procedure based on the sensitivity function. Despite this important difference, the result was rather similar to the variations found in DRM $\bm{C2}$. The variations of $f_1$ and $f_2$ consistently followed those of $\tilde{a}_1$ and $\tilde{a}_2$ but while $f_2$ remained quite constant for $\tilde{a}_1$, $f_1$ on the contrary had a strong parabolic variation for $\tilde{a}_2$. This corresponds well with the present quadratic term. Remarkably, this point was not discussed maybe because this in contradiction with the Schroeder's approach. This term cannot appear in his demonstration but it is inherent in the DRM rules \cite{Mrayati1988} which are recalled Table.~\ref{tab:tab4}. This shows that for increasing $f_2$, a constriction of $A_1$ (or $A_3$) leads to the increase of $f_1$ and similarly for decreasing $f_2$, a constriction of $A_2$ (or $A_4$) also leads to the increase of $f_1$. This is the relationship we retrieve quantitatively with the variations of $\tilde{a}_1$ and $\tilde{a}_2$. Moreover, this depends on the specific cut $\{\tfrac{L}{6},\tfrac{L}{2},\tfrac{5 L}{6}\}$ of the reduced DRM and any perturbation of it will change $(\tilde{a}_1,\tilde{a}_2)$ and invalidate the previous estimate.

\begin{figure}[ht]
\centering
\includegraphics[width=\reprintcolumnwidth]{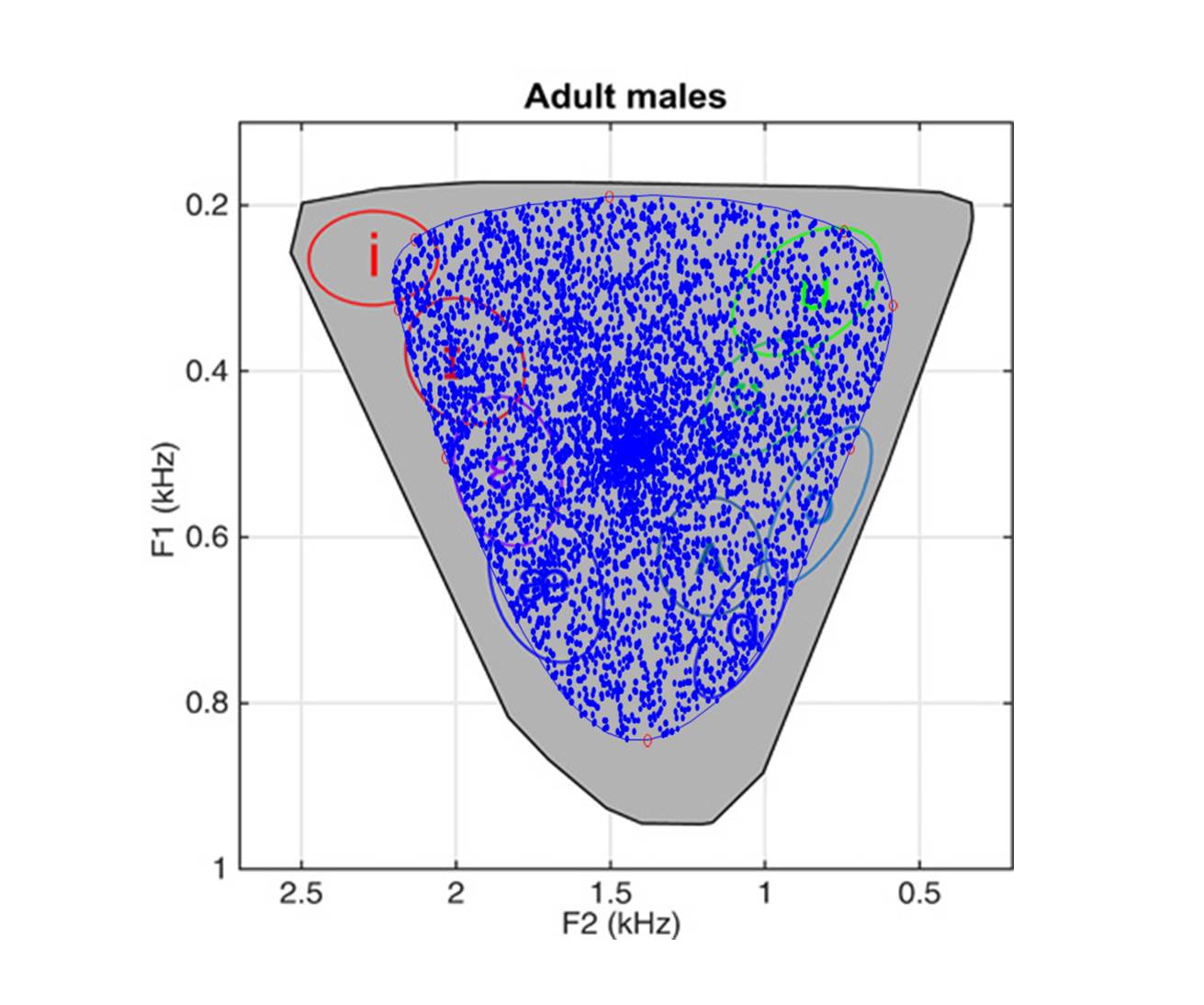}
\caption{\label{fig:FIG7}{Overlay of DRM $\bm{C2}$ data with \cite [Fig.~4 with $L=17.5\,cm$]{Boe2019}. This shows that a 4-tube model driven by 7 parameters has a significantly larger $(f_1, f_2)$ space than one with only 2 parameters. The human vowel space is depicted by background ellipses and only the \textipa{[i]} area is not well covered by this model.}}
\end{figure}

In the Fant model, the $X_c$ parameter controlling the length of the sections is involved, precluding the fit between $(\tilde{a}_1, \tilde{a}_2)$ and $(f_1, f_2)$. We conclude that implicit nonlinearities are difficult to analyze, emphasizing the value of having the simplest models in hand as a baseline. Moreover, even coefficients linked with symmetric configurations are enhanced in the Fant model and this is combined with length variations for explaining the many-to-one relationship observed in $\bm{C1}$. All these factors are present with real VT forms and this makes the existence of an analytical solution doubtful. In this context, the coordination function has the main property of generating for the non-optimal VT forms the regular relation existing with the generic model and the reduced DRM. In other words, the coordination function is an optimal way to exploit the articulatory potential of a vocal tract to create a vowel space as well as to generate a one-to-one correspondence. The residue of $\bm{C1}$ points (Fig.~\ref{fig:FIG4}b) with better contrast might be unrecoverable. Following this reasoning, most of the points produced with 7 parameters \cite[Fig.~4 with $L=17.5\,cm$]{Boe2019} and especially those located in the crown outside the DRM $\bm{C2}$ could be excluded from the definition of the articulatory potential of a closed-open tube. It appears in Fig.~\ref{fig:FIG7} that these points are outside the human vowel space, with the exception of \textipa{[i]} which is outside of the DRM $\bm{C2}$. This suggests that the third parameter we have discarded has a role for this vowel and is a call to extend the present analysis.

\section{\label{sec:6} Conclusion}
After \cite{Carre2004,Carre2009}, we showed that the vowel space can be described as a pre-existing mathematical object. After \cite{BoePerrier1990}, this has the disadvantage that it bears little resemblance to the human VT. To resolve this contradiction, a Platonic proposal is to consider that humans have discovered how to make vowels with their non-optimal vocal tract by using the coordination function. The first way of coding $\{\bm{\Omega},\bm{\Psi_1},\bm{\Psi_2}\}$ from $\{\bm{i},\bm{j},\bm{k}\}$ as in Table~\ref{tab:tab2} is abstract but the second Table~\ref{tab:tab3} is on the contrary adapted to think about this evolutionary process. Controlling $(\tilde{a}_1, \tilde{a}_2)$ is crucial because these determine the variations of $(f_1,f_2)$. The variations of the area function that we obtained with the generic function are not plausible and this problem persists even if the static data are from real measurements. With the articulatory parameters, the coordination function can indirectly vary $(\tilde{a}_1,\tilde{a}_2)$ to implement real deformations of the VT due to tongue movements. Because these latent variables represent the degree of VT antisymmetry, their control is potentially assumed by agonist-antagonist muscle relationships as well as with somatosensory feedback. Second, vowel production is geometrically constrained and is not just a control problem. This means that the acoustic contrast $df_i$ is determined by the volume $\left[\bm{\Omega}\pm\bm{\Psi_1}\right]$ which is indexed by $(\tilde{a}_1, \tilde{a}_2)$. For example, a restriction of the range of $X_c$ related to the ability to rotate the tongue in the mouth has a roughly proportional effect on $(\tilde{a}_1, \tilde{a}_2)$ and then on the size of the space $(f_1, f_2)$. For monkeys, this capacity is strongly reduced because they have a fixed larynx and a small pharyngeal cavity. The anatomical ranges observed by \cite{Fitch2016} are quite small and this provides a new objective criterion for judging whether they are truly "speech ready".

\docsection{Multimedia and supplementary material}
All simulations of this article are reproduced by the code of the companion
repository \repourl{} (validated Python translation and original Matlab
supplement).  The multimedia files are available in \texttt{multimedia/}:
\begin{itemize}\small
\multimedia{MM1.wav}{The 8 vowel sounds \textipa{[1, u, o, O, a, E, e, i]} corresponding to the characteristic phases for the generic model. This is a File of type "wav" (180 Ko).}\label{vowelsgeneric}
\multimedia{MM2.avi}{Variations of the area function from $0$ to $2\pi$ for the generic model. This is a File of type "avi" (145 Ko).}\label{areageneric}
\multimedia{MM3.avi}{Variations of the area function from $0$ to $2\pi$ for the DRM. This is a File of type "avi" (65 Ko).}\label{areaDRM}
\multimedia{MM4.wav}{The 8 vowel sounds corresponding to the characteristic phases for the DRM. This is a File of type "wav" (180 Ko).}\label{vowelsDRM}
\multimedia{MM5.avi}{Variations of the area function from $0$ to $2\pi$ for the Fant model. This is a File of type "avi" (63 Ko).}\label{areaFant}
\multimedia{MM6.wav}{The 8 vowel sounds corresponding to the characteristic phases for the Fant model. This is a File of type "wav" (180 Ko).}\label{vowelsFant}
\end{itemize}

\bibliographystyle{plainnat}
\bibliography{Paperbib}

\onecolumn
\docsection{Tables}

\noindent\begin{minipage}[t]{0.485\textwidth}
\centering
\small
\renewcommand{\arraystretch}{1.7}%
\begin{tabular}{ccccccc}
\hline\hline
Parameter &$\bm{i}$  &$\bm{j}$ &$\bm{k}$ &$\bm{\Omega}$ &$\bm{\Psi_1}$ &$\bm{\Psi_2}$ \\
\hline
$a_1$ & $1$ & $-2$ & $1$ & $0$ & $2$ & $0$ \\
$a_2$ & $1$ & $0$ & $-1$ & $0$ & $cos^{-1}\frac{\pi}{6}$ & $\frac{\pi}{2}$  \\
\hline\hline
\end{tabular}
\captionof{table}{\label{tab:tab1} Conversion of $\{\bm{i},\bm{j}, \bm{k}\}$ into $\{\bm{\Omega},\bm{\Psi_1},\bm{\Psi_2}\}$ for $(a_1,a_2)$ taken as parameters of the coordination function.}

\vspace{10pt}
\renewcommand{\arraystretch}{1.7}%
\setlength{\tabcolsep}{3pt}
\begin{tabular}{ccccc}
\hline\hline
Parameter &$x$ &$i(x)$  &$j(x)$ &$k(x)$ \\
\hline
$P_1$ & $0$ & $3$ & $-1$ & $1$ \\
$P_2$ & $L/3$ & $0.5$ & $0$ & $2.5$  \\
\hline\hline
Parameter & $x$ &$\Omega(x)$ &$\Psi_1(x)$ &$\Psi_2(x)$ \\
\hline
$P_1$ & $0$ & $1$ & $2\,cos^{-1}(atan (\frac{2}{3} \,sin\,\frac{\pi}{3}))$ & $atan(\frac{2}{3} \,sin\,\frac{\pi}{3})$ \\
$P_2$ & $L/3$ & $1$ & $cos^{-1}(atan (-\frac{4}{3} \,sin\,\frac{\pi}{3}))$ & $atan(-\frac{4}{3} \,sin\,\frac{\pi}{3})$  \\
\hline\hline
\end{tabular}
\captionof{table}{\label{tab:tab2} Setup of the coordination function (Eq.~\ref{Eq7}) for the 2 first parameters of the DRM controlling the areas of the two first sections from glottis and the two others are set in the text. The model is derived from $\{\bm{i},\bm{j},\bm{k}\}$ of the generic model at given points $x$. The reconstruction of the 4-tube $P(i,\rho,\theta)$ function is realized with $n=120$ tubelets having length $\tfrac{L}{n}$ with $L=17.5\, cm$. The proportion of $n$ composing each of the 4 sections is $\{\tfrac{n}{6},\tfrac{n}{3},\tfrac{n}{3},\tfrac{n}{6}\}$ and the value of $P(i,\rho,\theta)$ for each is $\{P_1,P_2,P_3,P_4\}$.}
\end{minipage}
\hfill
\noindent\begin{minipage}[t]{0.485\textwidth}
\centering
\small
\begin{tabular}{cccc}
\hline\hline
 Parameter &$\bm{\Omega}$  &$\bm{\Psi_1}$ &$\bm{\Psi_2}$ \\
\hline
$X_c$ & $L_c/2$ & $0.3\,(L_c-l)$ & $5\Pi/3$  \\
$A_c$ & $-1.5\,A$ & $2\,A$ & $\Pi$  \\
$A_l$ & $A/2$ & $-A$ & $\Pi/3$ \\
$L$ & $\overline{L}$ & $\Delta L$ & $\Pi/3$ \\
\hline\hline
\end{tabular}
\captionof{table}{\label{tab:tab3} Specification of $\{\bm{\Omega},\bm{\Psi_1},\bm{\Psi_2}\}$ of Eq.~\ref{Eq7} for the 4 parameters of the Fant model with the rules described in the text. For parameters $\{A_c,A_l,L\}$ the underlying geometrical parameters are: $A=1 \,cm^2$, $\overline{L}=17.5 \,cm$ and $\Delta L=1.5 \,cm$. The length of sections is defined in number of tubelets so that for $X_c$ the underlying geometrical parameters are proportions of $n=200$: $L_c=0.9$ and $l=0.3$. The reconstruction of the 4-tube $P(i,\rho,\theta)$ function is realized after rounding, merging and before applying the soft rectification.}

\vspace{10pt}
\begin{tabular}{cccccc}
\hline\hline
&Constriction &$A_1$  &$A_2$ &$A_3$ &$A_4$ \\
\hline
&$f_1$ &$\nearrow$ &$\nearrow$ &$\searrow$ &$\searrow$  \\
&$f_2$ &$\nearrow$ &$\searrow$ &$\nearrow$ &$\searrow$  \\
\hline\hline
\end{tabular}
\captionof{table}{\label{tab:tab4} Variations of formant frequencies $f_i$ according to a constriction of each of the four sections from glottis to lips. This reduction of area also corresponds to a variation of the cosine coefficients $\tilde{a}_i$ which is inverse to this of $f_i$.}
\end{minipage}

\end{document}